\documentclass{PoS}
\usepackage{adjustbox}
\usepackage[table]{xcolor}
\usepackage{makecell}
\title{A Search for IceCube Neutrinos from the First 33 Detected Gravitational Wave Events}

\ShortTitle{Neutrino Follow Up of Gravitational Wave Events}

\author{
The IceCube Collaboration\footnote{For collaboration list, see PoS(ICRC2019) 1177.}\\
{\itshape \href{http://icecube.wisc.edu/collaboration/authors/icrc19_icecube}{http://icecube.wisc.edu/collaboration/authors/icrc19\_icecube}}\\
E-mail: \email{rhussain@icecube.wisc.edu}
}

\abstract{
The discoveries of high-energy astrophysical neutrinos by IceCube in 2013 and of gravitational waves by LIGO in 2015 have enabled a new era of multi-messenger astronomy. Gravitational waves can identify the merging of compact objects such as neutron stars and black holes. These compact mergers, especially neutron star mergers, are potential neutrino sources. We present an analysis searching for neutrinos from gravitational wave sources reported by the LIGO Virgo Collaboration (LVC). We use a dedicated transient likelihood analysis combining IceCube events with source localizations provided by LVC as spatial priors. We report results for all gravitational wave events from the O1, O2, and O3 observing runs.\\

\vspace{4mm}
{\bfseries Corresponding authors:}
\speaker{Raamis Hussain}$^{1}$, Justin Vandenbroucke$^{1}$, Joshua Wood$^{2}$\\
{$^{1}$ \itshape University of Wisconsin - Madison}\\ {$^{2}$ \itshape IceCube Collaboration}\\

}

\FullConference{36th International Cosmic Ray Conference -ICRC2019-\\
		July 24th - August 1st, 2019\\
		Madison, WI, U.S.A.}

\begin{document}
\section{Introduction}\label{sec:info}
With the release of the first Gravitational Wave Transients Catalog \cite{LIGO:2018mvr} and the start of the O3 observing run of the LIGO-Virgo Collaboration (LVC), there have been 33 detected compact binary mergers as of July 28th, 2019. Of these 33 detections, 28 have been classified as binary black hole mergers (BBH), 3 are thought to be binary neutron star mergers (BNS), and 2 events may be of terrestrial origin \cite{graceDB}.

Although several multi-messenger campaigns to find electromagnetic counterparts to gravitational waves (GWs) have been conducted, only one successful multi-messenger detection has been made \cite{GBM:2017lvd}. The observation of GW170817, and the associated short gamma-ray burst GRB170817A, provided a wealth of knowledge about the physical processes and dynamics of the astrophysical system \cite{ANTARES:2017bia,Monitor:2017mdv}.

Neutrino follow up searches of gravitational wave events can provide information complementary to electromagnetic counterparts. For example, neutrinos can provide information about particle acceleration mechanisms \cite{Halzen:2002pg}, jet dynamics \cite{Razzaque:2003uv}, and the environment near the source \cite{Murase:2015xka}. In addition to providing insight into the physics of these sources, detection of neutrinos from GW sources can greatly improve the localization compared to detection of GWs alone. Searches for joint GW and high-energy neutrino events have produced no significant detections to date \cite{ANTARES:2017bia,Adrian-Martinez:2016xgn,Albert:2018jnn}.

IceCube is a cubic kilometer neutrino observatory located at the geographic South Pole. It has a duty cycle close to 100\% and is able to observe the full sky at all times \cite{Kintscher:2016uqh}. Rapid analysis of neutrino data from IceCube presents a unique opportunity to quickly identify joint sources of GWs and neutrinos and to report these findings to electromagnetic telescopes for further study.

In these proceedings, we present a comprehensive neutrino follow up to every detected GW event from the LVC O1, O2 and partially completed O3 observing run. The unbinned maximum likelihood used for the search is described in Section \ref{sec:methods}. The data used in this analysis are described in Section \ref{sec:data}. The results for each follow up are summarized in Section \ref{sec:results}. We conclude with a discussion of the results and future work in Section \ref{sec:discussion}.

\section{Method}\label{sec:methods}
We perform an unbinned maximum likelihood analysis which uses the LVC skymap as a spatial prior. The likelihood we use is
\begin{equation}
    \textit{L} = \frac{e^{-(n_{\mathrm{s}} + n_{\mathrm{b}})} (n_{\mathrm{s}} + n_{\mathrm{b}})^{N}}{N!} \prod_{i=1}^{N} \frac{n_{\mathrm{s}} \textit{S}_{i}(x_i,E_i;\gamma) + n_{\mathrm{b}} \textit{B}_{i}(x_i,E_i)}{n_{\mathrm{s}} + n_{\mathrm{b}}}
\end{equation}
where $N$ is the total number of observed neutrino events, $n_{\mathrm{b}}$ is the expected number of background events, $n_{\mathrm{s}}$ is the number of signal events, and $\gamma$ is the spectral index of the source. For an event with a given energy $E_i$ and direction $x_i$, we form a PDF of signal correlation \textit{$S_i$}($x_i$,$E_i$;$\gamma$) and a PDF of background correlation \textit{$B_i$}($x_i$,$E_i$). Both the signal and background PDFs consist of a spatial and energy component

\begin{equation}
    \textit{S}_i= S_{\mathrm{space}}(x_i,\sigma_i|x_{\mathrm{s}}) \cdot S_{\mathrm{energy}}(x_i,E_i|\gamma)
    ;\quad
    \textit{B}_i = B_{\mathrm{space}}(x_i) \cdot B_{\mathrm{energy}}(x_i,E_i)
\end{equation}
\\
We divide the sky into equal-area bins using HEALPix \cite{healpix}. The pixels are roughly 0.01 deg$^2$. In the equations above, the location of the pixel being tested is $x_{\mathrm{s}}$, while $x_i$ and $\sigma_i$ are the reconstructed direction and estimated angular uncertainty of each neutrino event. 

The test statistic is the log of the likelihood ratio

\begin{equation}
    \mathrm{TS} = 2 \ln\left(\frac{\textit{L}(\hat{n_{\mathrm{s}}},\hat{\gamma})}{\textit{L}(n_{\mathrm{s}}=0)} \right)
    ;\quad
    w = \frac{P_{\mathrm{GW}}(x_{\mathrm{s}})}{\Omega_{\mathrm{pixel}}}
\end{equation}
where $\hat{n_{\mathrm{s}}}$ and $\hat{\gamma}$ are the free parameters in the maximum likelihood fit.

We incorporate the GW spatial prior by defining a weight, shown above, for every pixel in the skymap. This weight represents the probability density of the GW source being at a given position in the sky. The weight is then normalized and multiplied by the signal likelihood which results in a modified test statistic

\begin{equation}
    \Lambda = 2 \ln\left(\frac{\textit{L}(\hat{n_{\mathrm{s}}},\hat{\gamma}) \cdot w}{\textit{L}(n_{\mathrm{s}}=0)} \right) = \mathrm{TS} + 2\ln(w) = 2\left(-n_{\mathrm{s}} + \sum_{i=1}^{N} \ln\left[1 + \frac{n_{\mathrm{s}} \textit{S}_{i}}{n_{\mathrm{b}} \textit{B}_{i}}\right]\right) + 2\ln(w)
\end{equation}

To test for a GW+neutrino coincidence, we consider a $\pm$500~s time window centered on the GW event time. This is a conservative time window derived by considering a range of prompt neutrino emission mechanisms from gamma-ray bursts \cite{Baret:2011tk}. We perform a scan over the  full sky and maximize the likelihood ratio with respect to $n_{\mathrm{s}}$ and $\gamma$ at every pixel. The largest test statistic in the sky is considered the best fit position for that scan.
    
\begin{figure}
    \centering
    \includegraphics[width=0.8\textwidth]{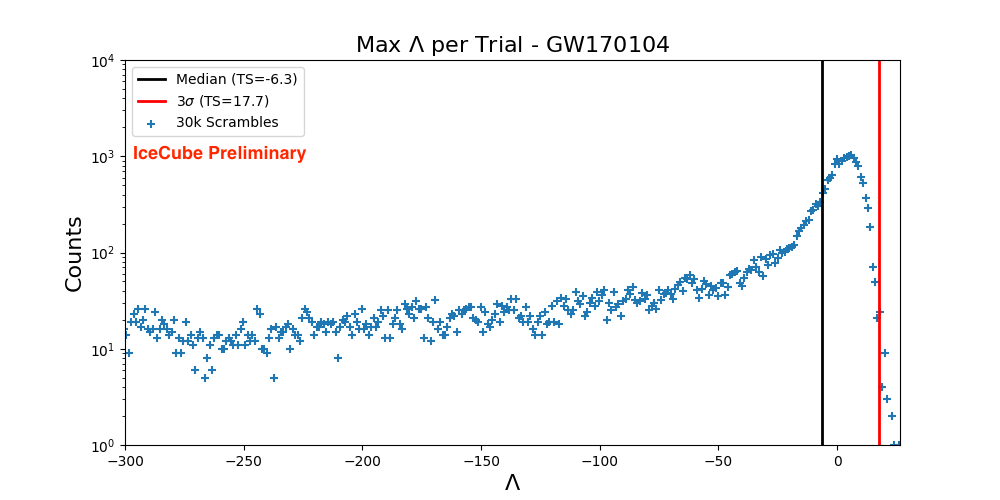}
    \caption{Background TS distribution for GW170104. The distribution is constructed by taking the maximum $\Lambda$ per trial for 30k background-only trial. More than half of the trials return negative $\Lambda$ values since most of the pixels in the sky have a very low GW spatial prior weight, $w$, from the GW PDF and thus are heavily penalized.}
    \label{fig:bkgTS}
\end{figure}

To calculate the significance of a given observation, we compute a p-value by comparing our observed $\Lambda$ to a background distribution which is built from 30,000 trials using randomized neutrinos and a fixed gravitational wave skymap for each trial. Neutrinos are randomized by scrambling their arrival time and recomputing their direction based on their new randomly assigned time. All trials use the same GW skymap. An example background distribution is shown in Figure \ref{fig:bkgTS}. This p-value quantifies the significance of the observed neutrinos being associated with a point-like source located within the given GW skymap. 

\begin{figure}
    \centering
    \includegraphics[width=0.55\textwidth]{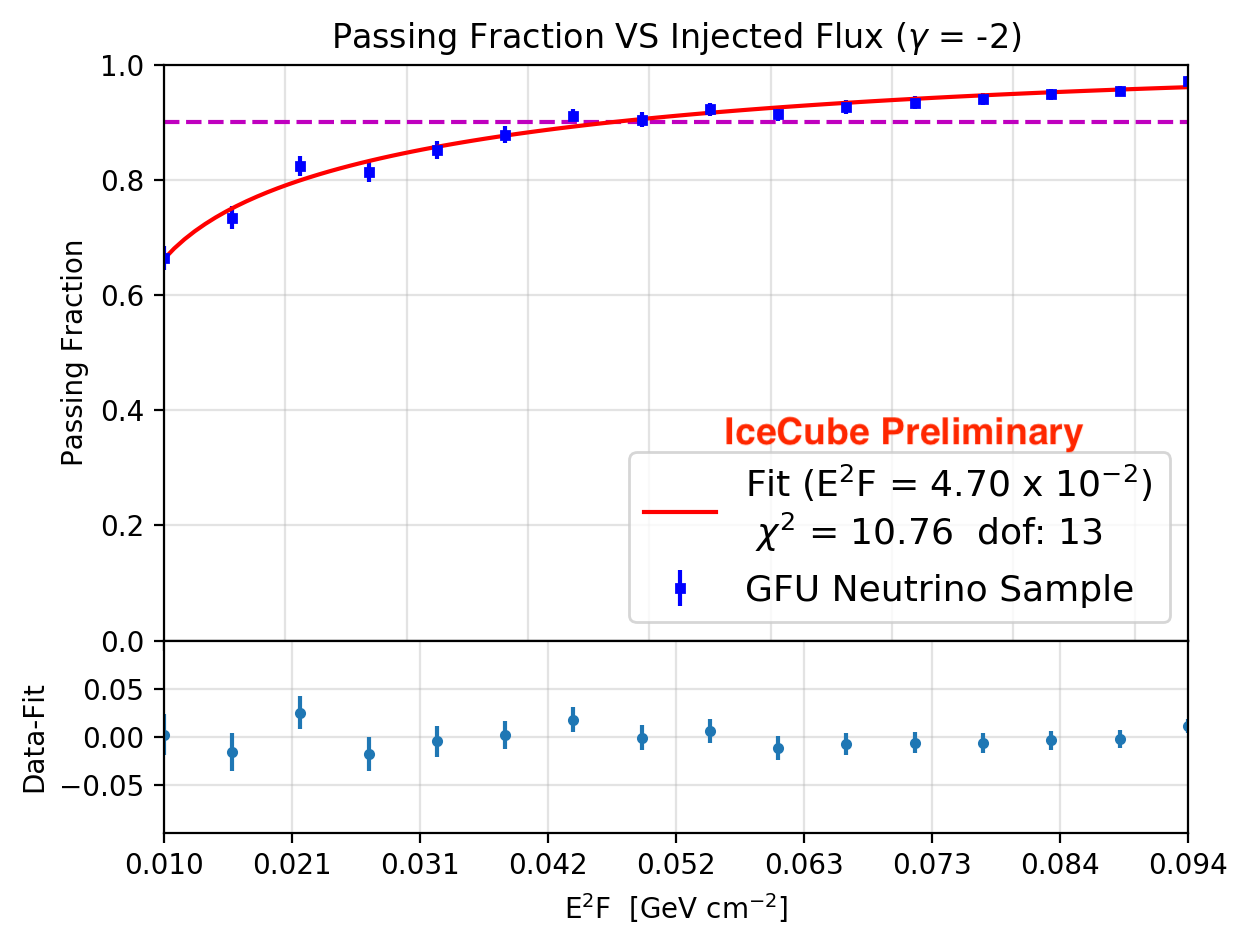}
    \caption{90\% Sensitivity flux for GW170104, computed by injecting an increasing neutrino flux according to an $E^{-2}$ power law spectrum and calculating the fraction of trials which return an observed $\Lambda$ greater than the median of the background distribution shown in Figure \ref{fig:bkgTS}. We use a $\chi^2$ cumulative distribution function (CDF) to fit the data and compute the flux which gives a 90\% passing fraction. This is defined as the sensitivity for the particular GW event.}
    \label{fig:sensitivity}
\end{figure}

To compute our sensitivity for each GW, we perform signal injection trials and compute the flux required such that 90\% of trials return an observed $\Lambda$ greater than the median $\Lambda$ of the background distribution. Signal neutrinos are selected from Monte Carlo and injected according to an $E^{-2}$ power law spectrum. An example of this procedure is shown in Figure \ref{fig:sensitivity}.

\section{Data Sample}\label{sec:data}
Neutrino data used in this analysis come from the IceCube Gamma-ray Follow Up (GFU) dataset, which is a sample of through-going muon tracks used for realtime analyses \cite{Kintscher:2016uqh}. The sample has a 6.7 mHz all-sky event rate with the dominant background in the Southern and Northern Hemispheres coming from atmospheric muons and atmospheric neutrinos, respectively. The sample has a median angular resolution of $\lesssim$1 deg for neutrino energies above 1 TeV \cite{Aartsen:2016lmt}.

\section{Results}\label{sec:results}
Table \ref{tab:results} summarizes the final results for each GW event. No significant neutrino correlation was found. We report 90\% C.L. upper limits (U.L.s) on the time-integrated neutrino flux scaled by event energy, $E^2F$, from each GW event. We also report 90\% U.L.s on the isotropic equivalent energy emitted in neutrinos within the $\pm500$~s time window, assuming the median value of the reported luminosity distance to the GW source.

The flux U.L.s are computed using the same procedure as the 90\% sensitivity flux. The difference between the U.L. and the sensitivity is that the U.L. uses the observed $\Lambda$ as the threshold rather than the median $\Lambda$ of the background. In cases where the observed $\Lambda$ is less than the median, we use the median $\Lambda$ as our threshold to be conservative.

We also report U.L.s on the isotropic equivalent energy, $E_{\mathrm{iso}}$, emitted in neutrinos over the 1000~s time window for each GW. This limit is derived using a relation between $E_{\mathrm{iso}}$ and the mean number of expected events at IceCube, taking into account the effective area of the detector and assuming an E$^{-2}$ powerlaw spectrum. We also marginalize over the 3D position of the source using skymaps and distance PDFs provided by LVC. $E_{\mathrm{iso}}$ is tuned such that the mean number of expected neutrino events at IceCube is 2.3, which corresponds to the 90\% C.L. Poisson U.L. given a non-observation of neutrinos at IceCube.

\begin{table}[htb!]
\centering
\resizebox{0.86\textwidth}{!}{
\rowcolors{2}{gray!25}{white}
\begin{tabular}{|c|c|c|c|c|c|c|c|c|}
 \hline
 \multicolumn{9}{|c|}{GW Event List} \\
 \hline
 Event & Type  & $\Omega$ ($\mathrm{deg}^2$) & FAR ($\mathrm{yr}^{-1}$) & p-Value & \makecell{90\% U.L. \\ (GeVcm$^{-2}$)} & \makecell{90\% U.L.$_{\mathrm{min}}$ \\(GeVcm$^{-2}$)} & \makecell{90\% U.L.$_{\mathrm{max}}$ \\(GeVcm$^{-2}$)} & $E_{\mathrm{iso}}$ (ergs)\\
 \hline
 GW150914   & BBH  & 180   & \textless1.00 x 10$^{-7}$ &  0.62 &  0.432  & 0.0296 & 1.03   & 1.60 x 10$^{52}$  \\ \hline 
 GW151012   & BBH  & 1555  & 7.92 x 10$^{-3}$          &  0.71 &  0.177  & 0.0286 & 0.821  & 8.14 x 10$^{52}$  \\ \hline 
 GW151226   & BBH  & 1033  & \textless1.00 x 10$^{-7}$ &  0.79 &  0.205  & 0.0286 & 0.904  & 1.45 x 10$^{52}$  \\ \hline 
 GW170104   & BBH  & 924   & \textless1.00 x 10$^{-7}$ &  0.54 &  0.0440 & 0.0286 & 0.667  & 7.11 x 10$^{52}$  \\ \hline
 GW170608   & BBH  & 396   & \textless1.00 x 10$^{-7}$ &  0.61 &  0.0365 & 0.0309 & 0.0821 & 9.40 x 10$^{51}$  \\ \hline 
 GW170729   & BBH  & 1033  & 1.80 x 10$^{-1}$          &  0.21 &  0.620  & 0.0286 & 1.02   & 5.39 x 10$^{53}$  \\ \hline 
 GW170809   & BBH  & 340   & \textless1.00 x 10$^{-7}$ &  0.60 &  0.270  & 0.0568 & 0.758  & 8.21 x 10$^{52}$  \\ \hline 
 GW170814   & BBH  & 87    & \textless1.00 x 10$^{-7}$ &  1.0  &  0.449  & 0.488  & 0.711  & 2.94 x 10$^{52}$  \\ \hline
 GW170817   & BNS  & 16    & \textless1.00 x 10$^{-7}$ &  0.19 &  0.274  & 0.180  & 0.429  & 1.37 x 10$^{50}$  \\ \hline 
 GW170818   & BBH  & 39    & 4.20 x 10$^{-5}$          &  0.52 &  0.0276 & 0.0364 & 0.0431 & 9.04 x 10$^{52}$  \\ \hline
 GW170823   & BBH  & 1651  & \textless1.00 x 10$^{-7}$ &  0.75 &  0.182  & 0.0286 & 0.796  & 2.46 x 10$^{53}$  \\ \hline 
 S190408an  & BBH  & 387   & \textless1.00 x 10$^{-7}$ &  0.13 &  0.0625 & 0.0337 & 0.606  & 1.81 x 10$^{53}$  \\ \hline 
 S190412m   & BBH  & 156   & \textless1.00 x 10$^{-7}$ &  0.18 &  0.0423 & 0.0286 & 0.048  & 5.39 x 10$^{52}$  \\ \hline 
 S190421ar  & BBH  & 1444  & 4.70x 10$^{-1}$           &  0.79 &  0.652  & 0.0420 & 1.15   & 1.65 x 10$^{53}$  \\ \hline 
 S190425z   & BNS  & 7461  & 1.43 x 10$^{-5}$          &  0.87 &  0.383  & 0.0286 & 1.06   & 1.90 x 10$^{51}$  \\ \hline 
 S190426c   & BNS  & 1131  & 6.14 x 10$^{-1}$          &  0.12 &  0.0685 & 0.0286 & 0.583  & 1.10 x 10$^{52}$  \\ \hline 
 S190503bf  & BBH  & 448   & 5.16 x 10$^{-2}$          &  0.49 &  0.581  & 0.227  & 0.821  & 1.43 x 10$^{52}$  \\ \hline 
 S190510g   & Ter  & 1166  & 2.79 x 10$^{-1}$          &  0.86 &  0.401  & 0.0286 & 0.610  & 2.76 x 10$^{51}$  \\ \hline 
 S190512at  & BBH  & 252   & 6.00 x 10$^{-2}$          &  0.84 &  0.341  & 0.0286 & 0.568  & 1.51 x 10$^{53}$  \\ \hline 
 S190513bm  & BBH  & 691   & 1.18 x 10$^{-5}$          &  1.0  &  0.187  & 0.0286 & 0.505  & 3.16 x 10$^{53}$  \\ \hline 
 S190517h   & BBH  & 939   & 7.49 x 10$^{-2}$          &  0.21 &  0.613  & 0.0286 & 1.06   & 5.78 x 10$^{53}$  \\ \hline 
 S190519bj  & BBH  & 967   & 1.80  x 10$^{-1}$         &  0.45 &  0.108  & 0.0286 & 0.639  & 8.15 x 10$^{53}$  \\ \hline
 S190521g   & BBH  & 765   & 1.20  x 10$^{-1}$         &  0.61 &  0.538  & 0.0391 & 0.966  & 1.14 x 10$^{54}$  \\ \hline 
 S190521r   & BBH  & 488   & 1.00 x 10$^{-2}$          &  0.095&  0.0654 & 0.0286 & 0.456  & 1.07 x 10$^{53}$  \\ \hline 
 S190602aq  & BBH  & 1172  & \textless1.00 x 10$^{-7}$ &  0.15 &  0.344  & 0.0286 & 0.732  & 4.84 x 10$^{52}$  \\ \hline 
 S190630ag  & BBH  & 1483  & 4.53 x 10$^{-6}$          &  0.63 &  0.307  & 0.0286 & 0.977  & 6.73 x 10$^{52}$  \\ \hline 
 S190701ah  & BBH  & 67    & 6.04 x 10$^{-1}$          &  1.0  &  0.0530 & 0.0286 & 0.176  & 2.81 x 10$^{53}$  \\ \hline 
 S190706ai  & BBH  & 1100  & 6.00 x 10$^{-2}$          &  1.0  &  0.199  & 0.0350 & 0.881  & 2.09 x 10$^{54}$  \\ \hline
 S190707q   & BBH  & 1375  & 1.66 x 10$^{-4}$          &  0.55 &  0.334  & 0.0286 & 0.763  & 4.99 x 10$^{52}$  \\ \hline
 S190718y   & Ter  & 7246  & 1.15                      &  0.67 &  0.135  & 0.0286 & 1.15   & 1.42 x 10$^{51}$  \\ \hline
 S190720a   & BBH  & 1559  & 0.120                     &  0.96 &  0.358  & 0.0286 & 1.08   & 6.29 x 10$^{52}$  \\ \hline
 S190727h   & BBH  & 841   & 4.34 x 10$^{-3}$          &  0.95 &  0.592  & 0.0350 & 0.983  & 6.98 x 10$^{53}$  \\ \hline
 S190728q   & BBH  & 104   & \textless1.00 x 10$^{-7}$ &  0.014&  0.0520 & 0.0295 & 0.0404 & 6.83 x 10$^{52}$  \\ \hline

\end{tabular}}
\caption{Results for every detected GW from the O1, O2, and O3 observing runs as of July 28th, 2019. Here $\Omega$ is the area of the 90\% containment region of the GW and FAR is the false alarm rate. These values are taken from GWTC-1 and the online event database GraceDB \cite{LIGO:2018mvr,graceDB}. We also report 90\% C.L. U.L.s using the best-fit value from the $\chi^2$ fit described in Figure \ref{fig:sensitivity}, as well as the minimum and maximum U.L.s assuming a point source hypothesis within the 90\% containment region of the GW skymap. $E_{\mathrm{iso}}$ is the U.L. on the isotropic equivalent energy emitted in neutrinos during 1000~s. BBH = Binary Black Hole, BNS = Binary Neutron Star, Ter = Terrestrial. }
\label{tab:results}
\end{table}

\newpage

\section{Conclusion}\label{sec:discussion}
We searched for neutrino emission within $\pm$500~s of all reported GW events from the O1, O2, and O3 observing runs as of July 28th, 2019. No significant correlations with neutrino events were found and, for each GW event, we placed 90\% C.L. U.L.s on the time-integrated neutrino flux and 90\% C.L. U.L.s on $E_{\mathrm{iso}}$.
In addition to performing the neutrino follow up, we have reported our results through GCN circulars for other observatories. An independent search for joint GW+neutrino events which evaluates a Bayesian odds ratio of a detected signal is presented in \cite{llama:2019icrc}. The Bayesian approach produces results similar to those presented in this work. Results from both analyses are sent in IceCube's GCN circulars. Generally these results are sent out within an hour of the initial GCN notice sent by LVC. In the future we plan to send rapid GCN notices to inform the community of any significant neutrinos that should be followed up by electromagnetic observatories. 

During the current O3 observing run, 22 binary mergers have been detected thus far. The estimated rates before the start of the O3 run were $\sim$1/week for BBH mergers and $\sim$1/month for BNS mergers. So far the observed rates have been consistent with these expectations. Assuming this rate remains stable, we should expect to see roughly 40 more GW events by the end of the O3 run. With an increasing sample of GWs, we plan on performing population analyses on BBH mergers and BNS mergers to examine the individual source classes as potential neutrino emitters. In future work, we also plan to extend this analysis by searching for neutrino emission on timescales longer than $\pm500$~s. For BNS mergers especially, searches over longer timescales will have a higher chance of finding neutrino correlations if significant neutrino production occurs during the kilonova phase following the BNS merger \cite{Kimura:2018vvz,Fang:2017tla}.

\begin{center}
\begin{tabular}{ccc}
    \includegraphics[width=0.3\textwidth]{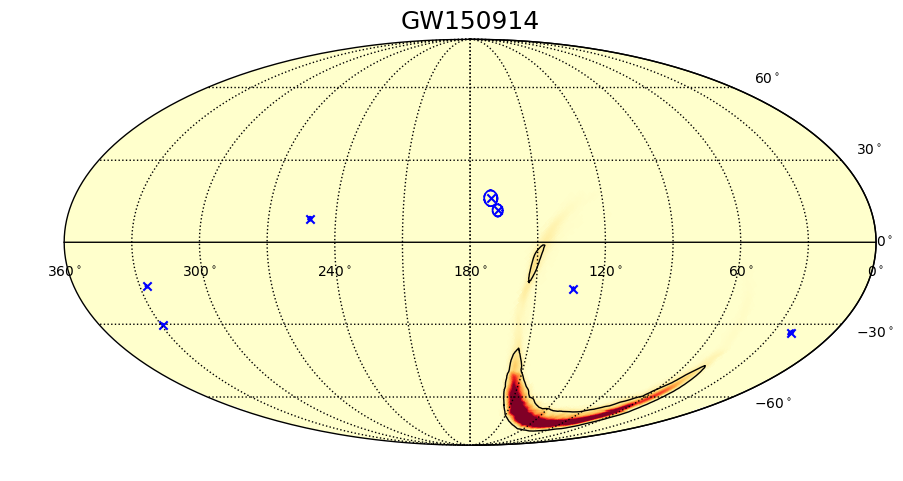} &
    \includegraphics[width=0.3\textwidth]{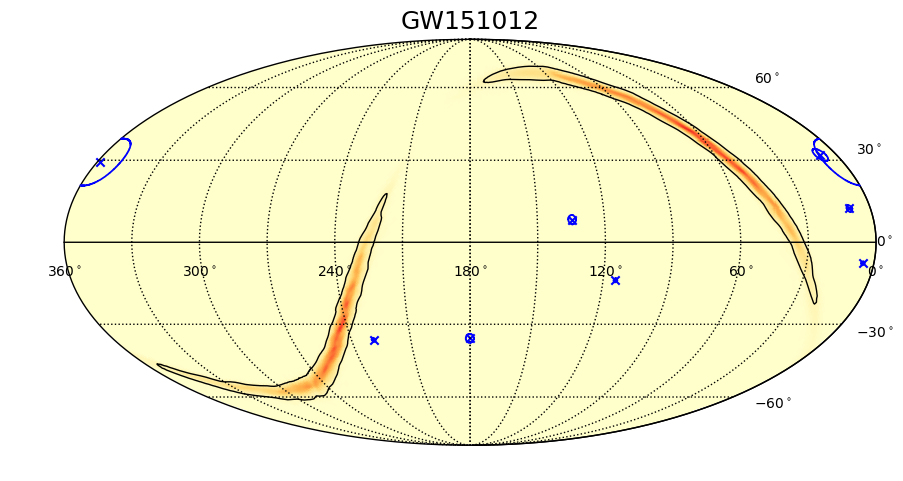} &
    \includegraphics[width=0.3\textwidth]{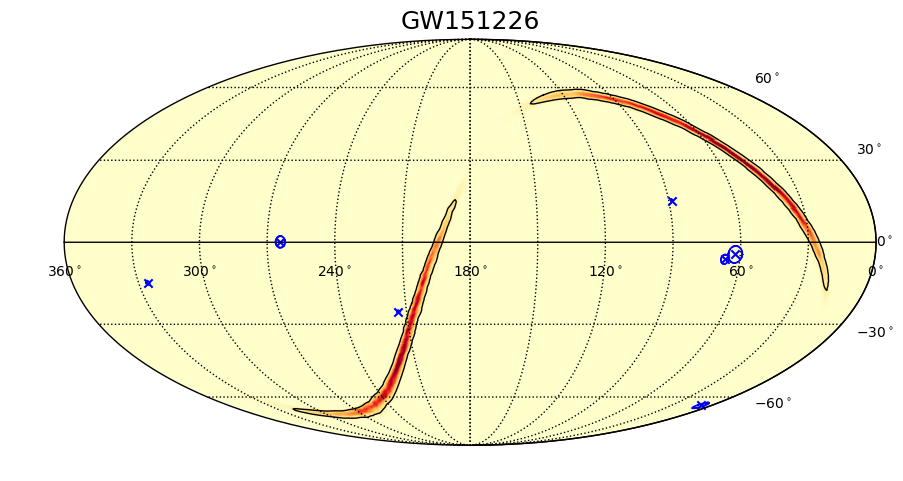} \\
    \includegraphics[width=0.3\textwidth]{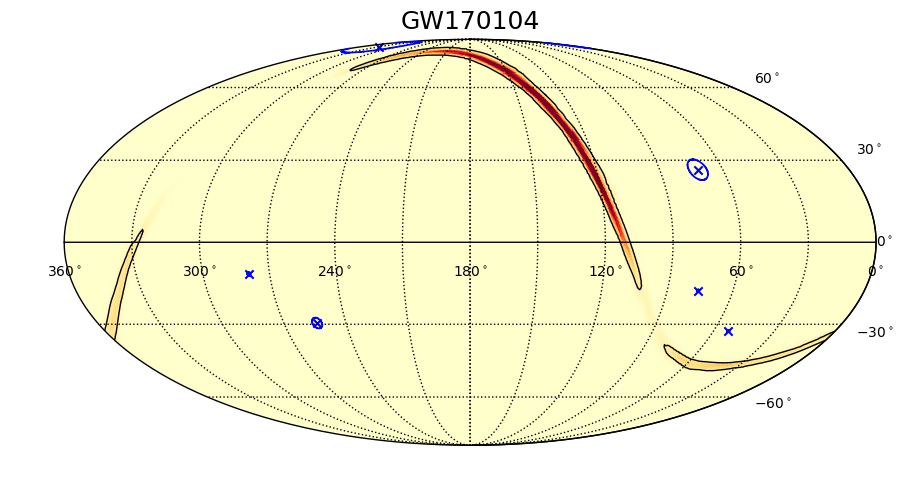} &
    \includegraphics[width=0.3\textwidth]{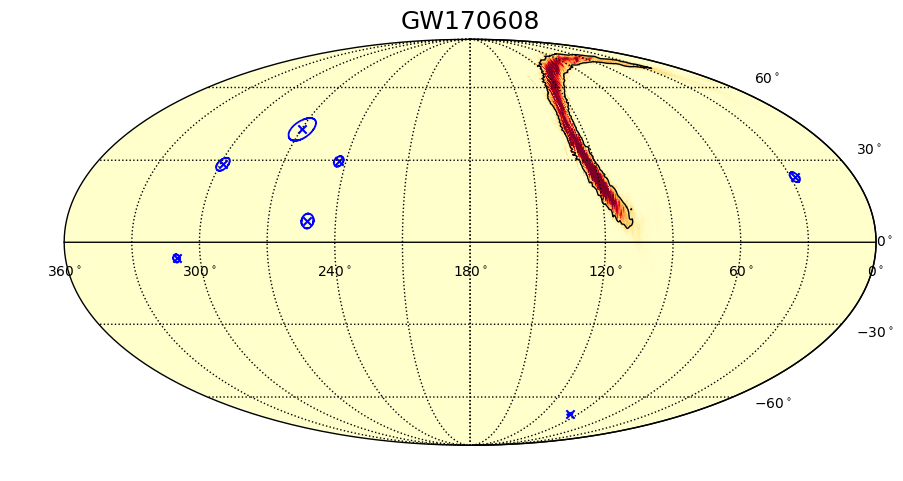} &
    \includegraphics[width=0.3\textwidth]{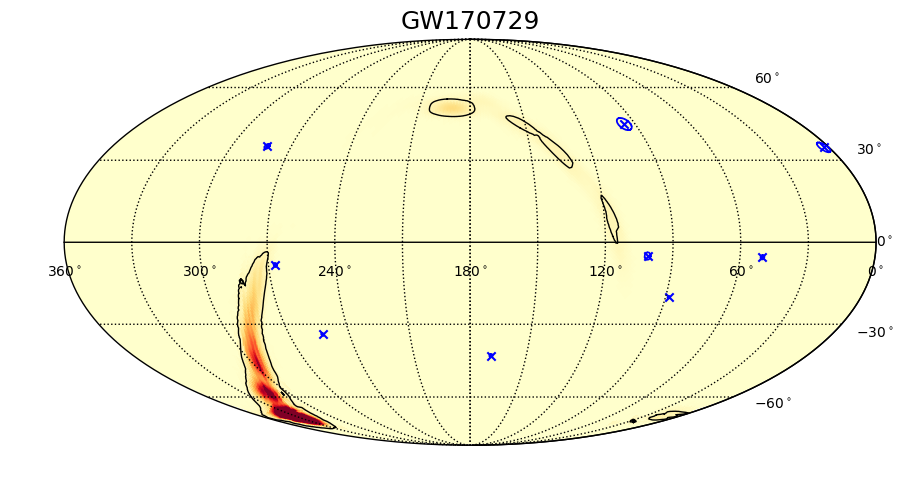} \\
    \includegraphics[width=0.3\textwidth]{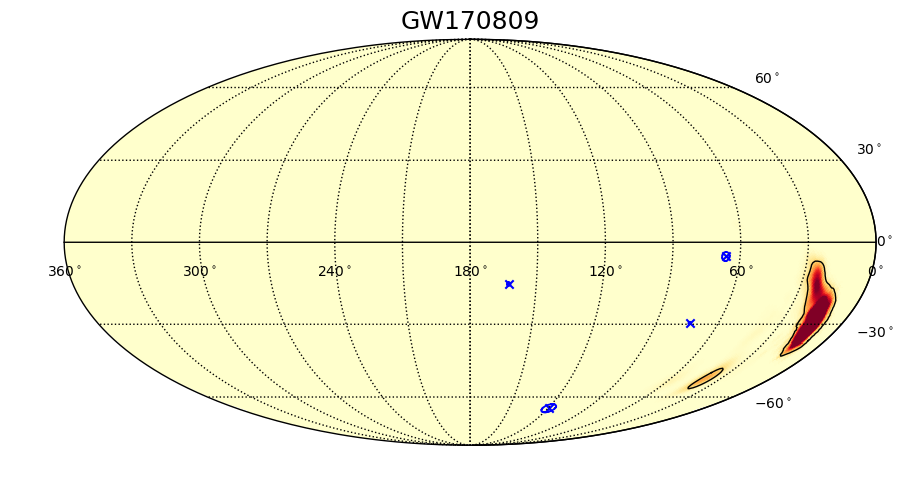} &
    \includegraphics[width=0.3\textwidth]{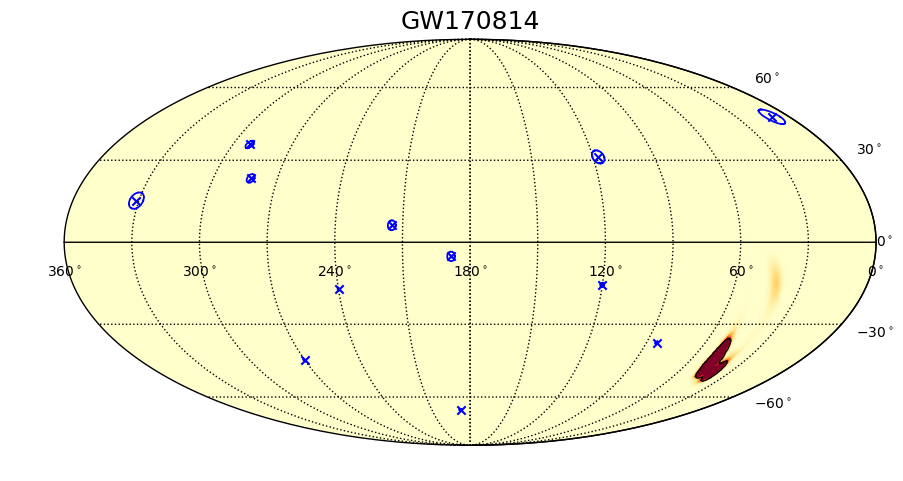} &
    \includegraphics[width=0.3\textwidth]{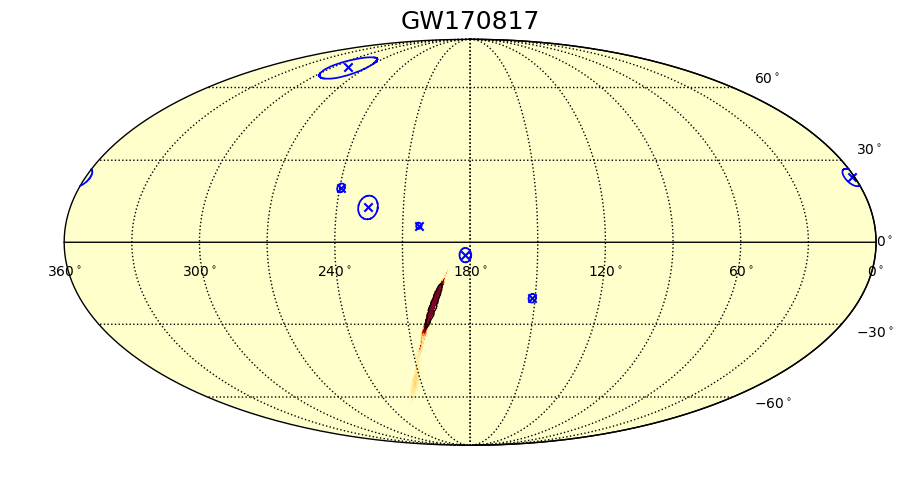} \\
    \includegraphics[width=0.3\textwidth]{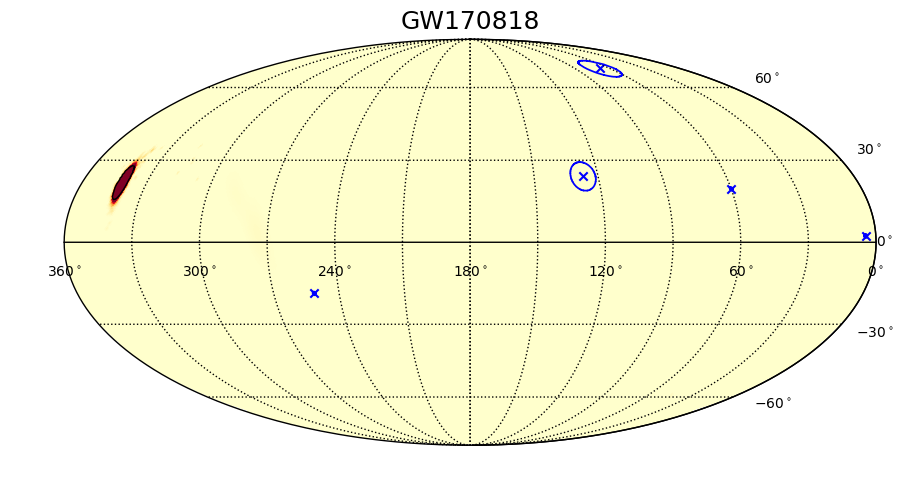} &
    \includegraphics[width=0.3\textwidth]{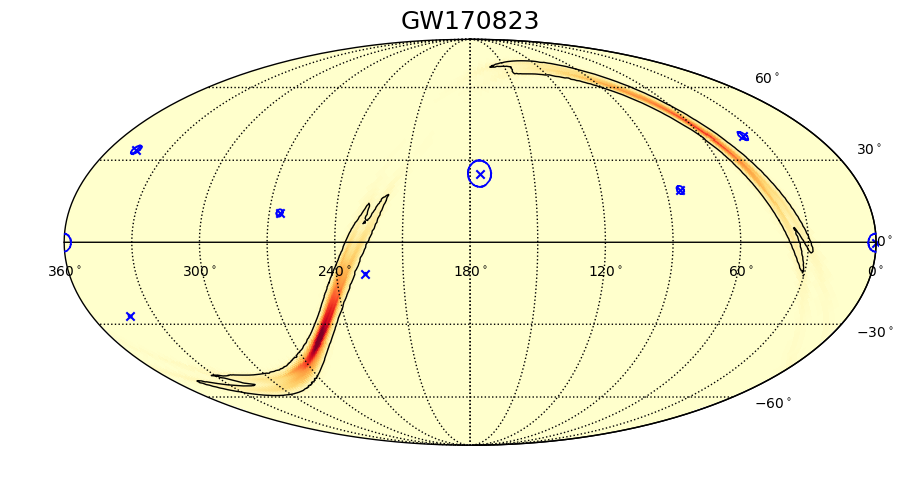} &
    \includegraphics[width=0.3\textwidth]{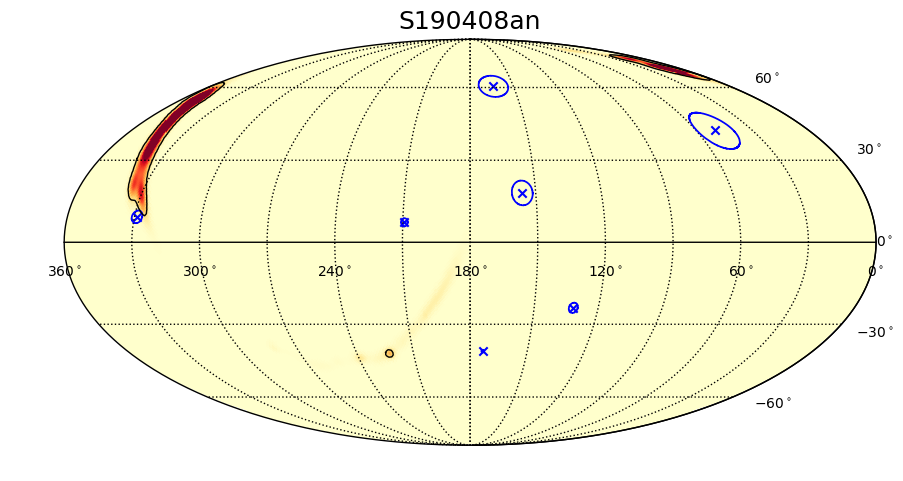} \\
    \multicolumn{3}{c}{\color{red}{{\Large{\textbf{IceCube Preliminary}}}}}
\end{tabular}
\end{center}

\newpage
\begin{figure}[htb!]
\begin{center}
\begin{tabular}{ccc}
    \includegraphics[width=0.3\textwidth]{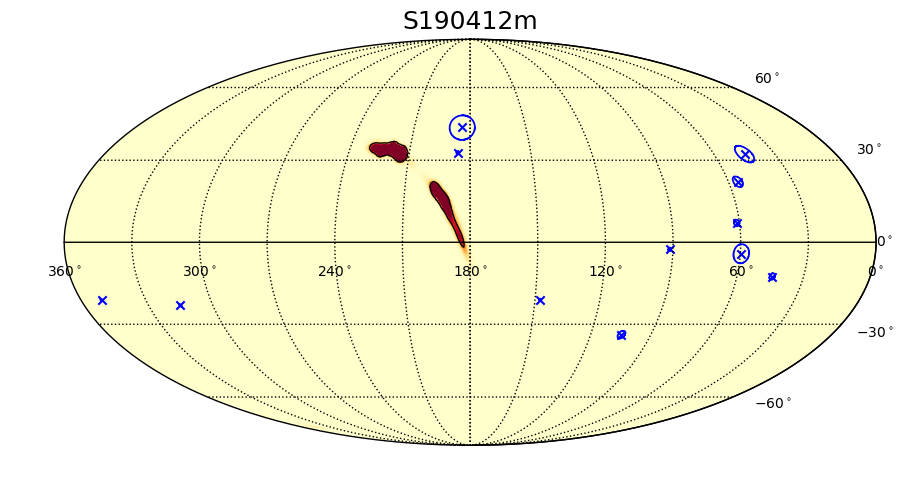} &
    \includegraphics[width=0.3\textwidth]{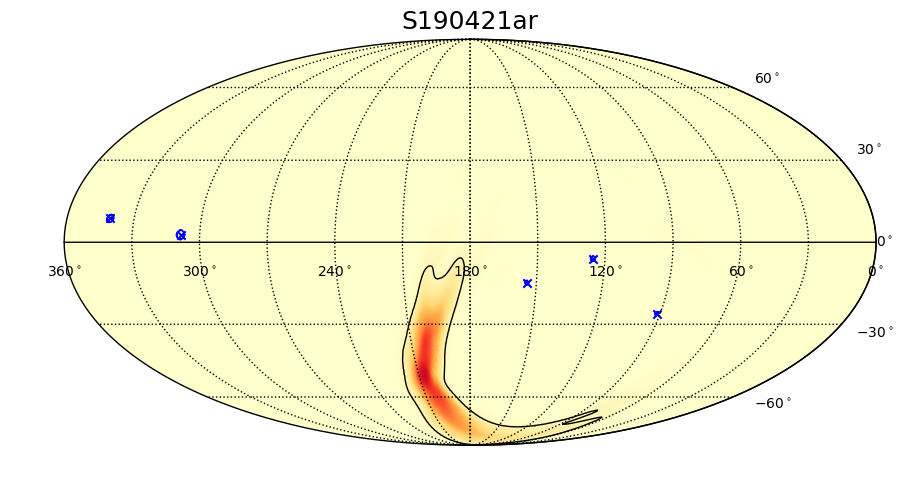} &
    \includegraphics[width=0.3\textwidth]{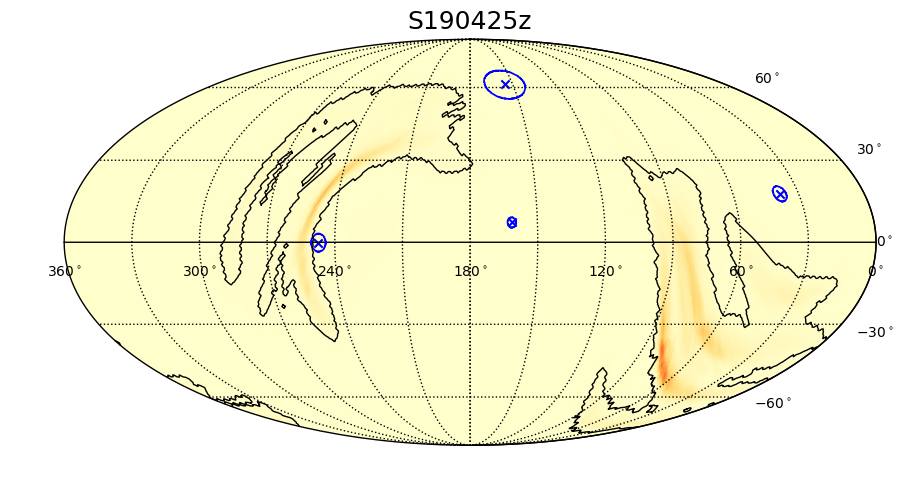} \\
    \includegraphics[width=0.3\textwidth]{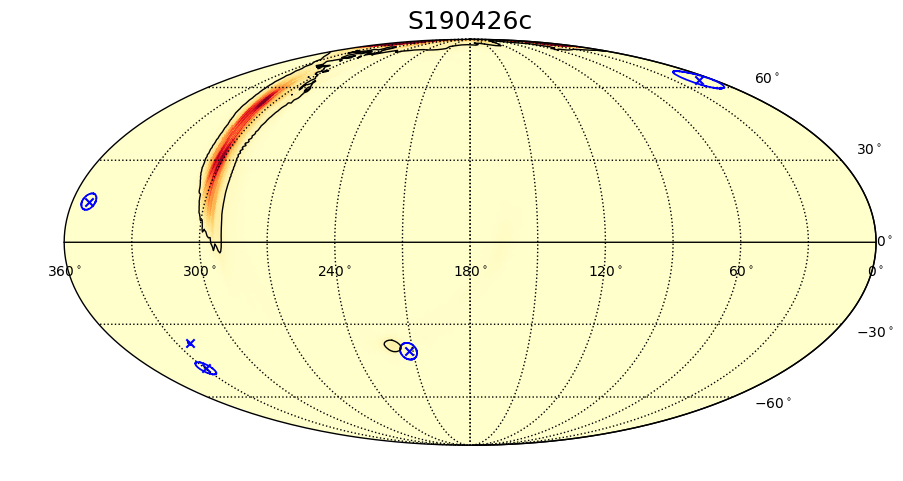} &
    \includegraphics[width=0.3\textwidth]{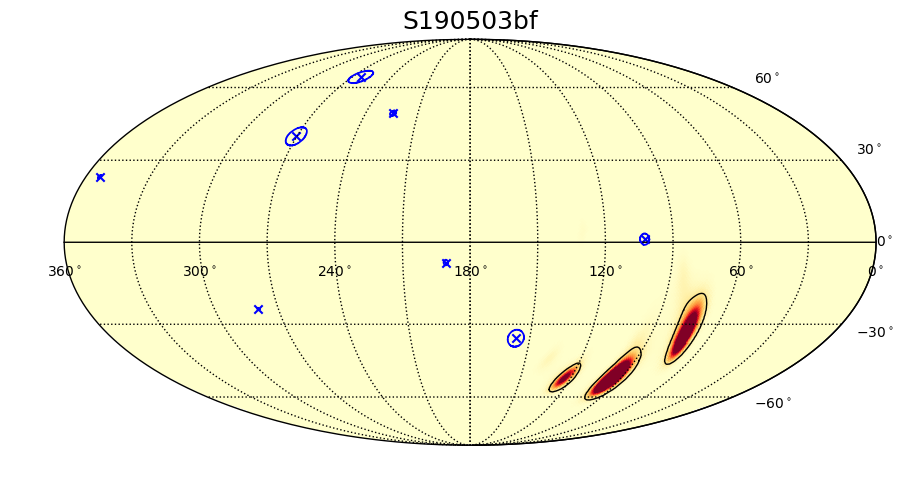} &
    \includegraphics[width=0.3\textwidth]{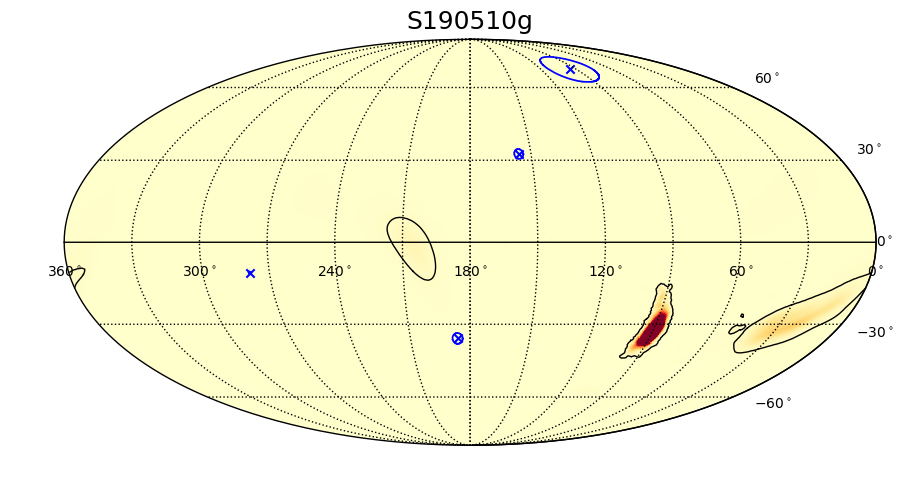} \\
    \includegraphics[width=0.3\textwidth]{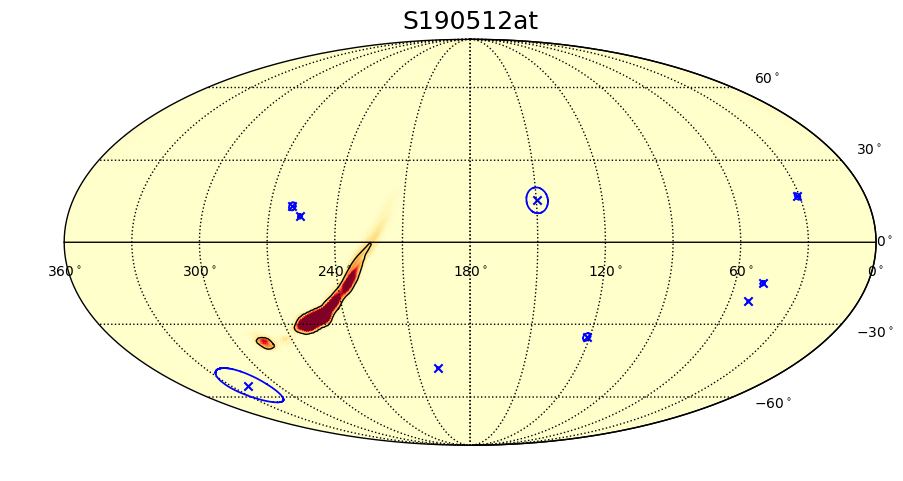} &
    \includegraphics[width=0.3\linewidth]{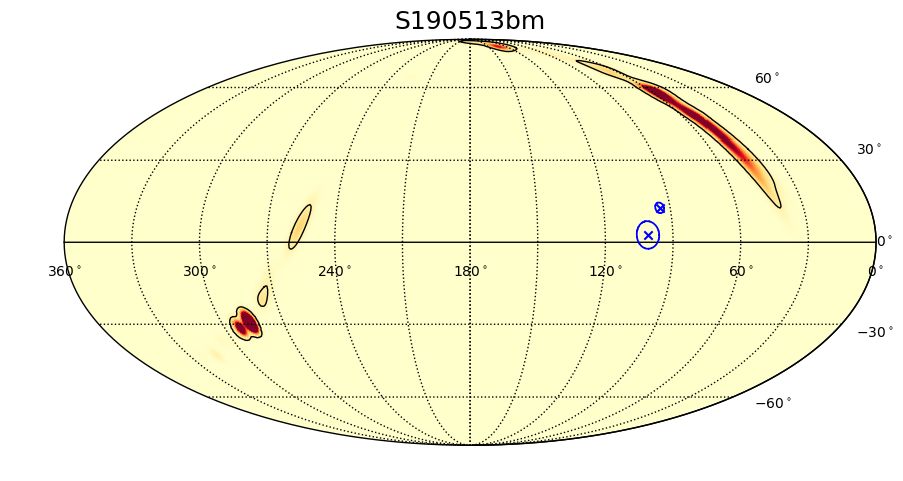} &
    \includegraphics[width=0.3\linewidth]{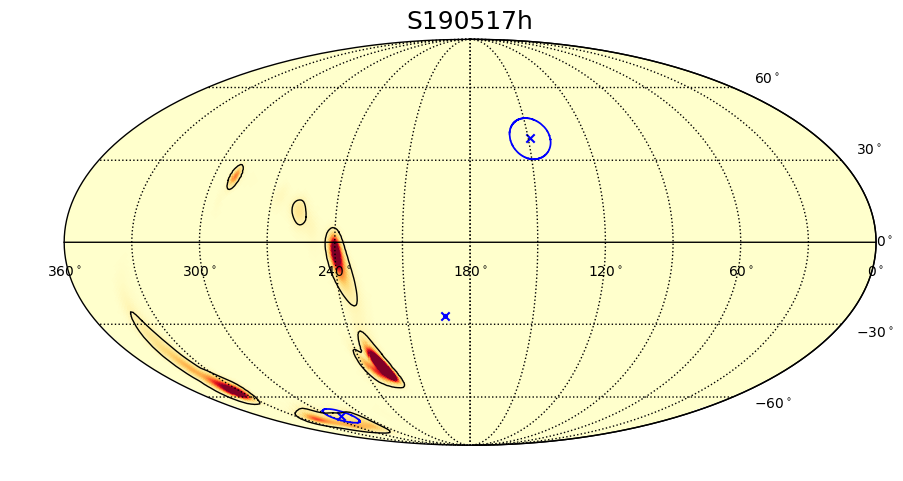} \\
    \includegraphics[width=0.3\textwidth]{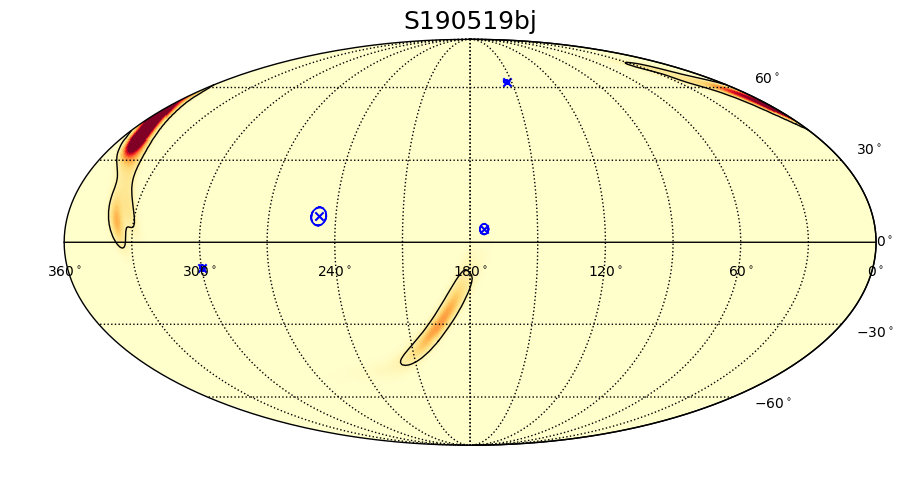} &
    \includegraphics[width=0.3\textwidth]{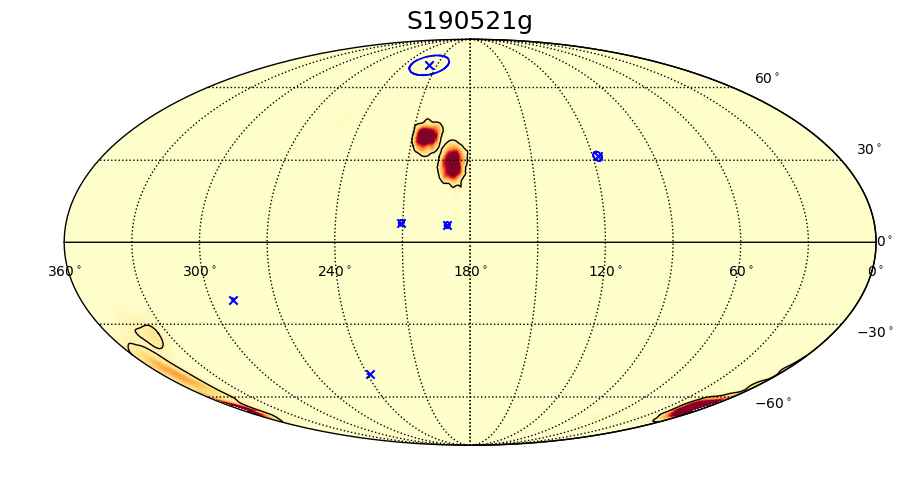} &
    \includegraphics[width=0.3\textwidth]{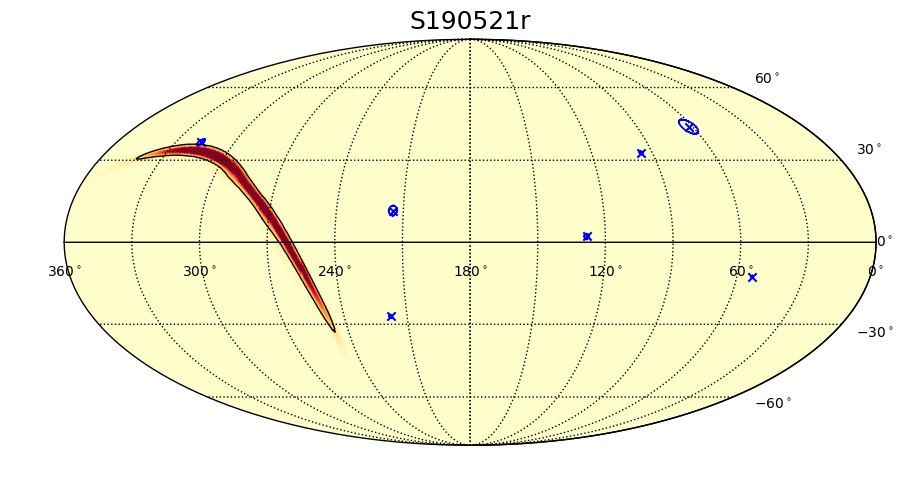} \\
    \includegraphics[width=0.3\textwidth]{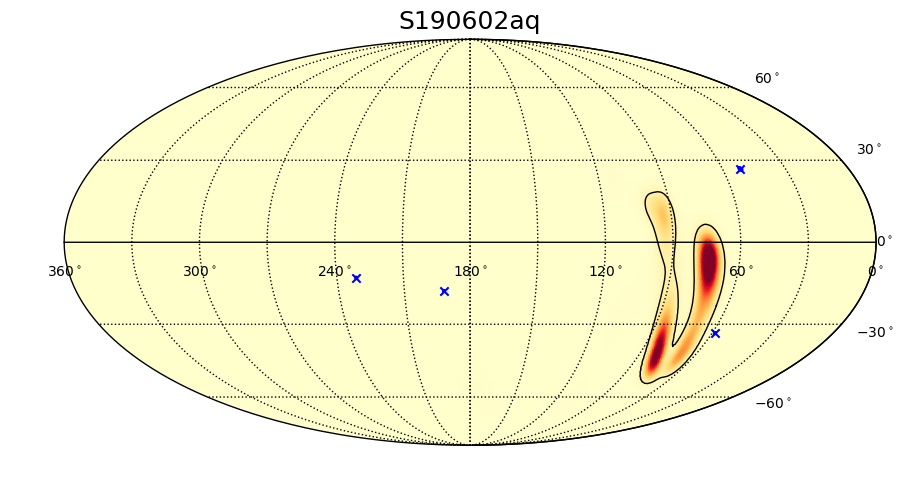} &
    \includegraphics[width=0.3\textwidth]{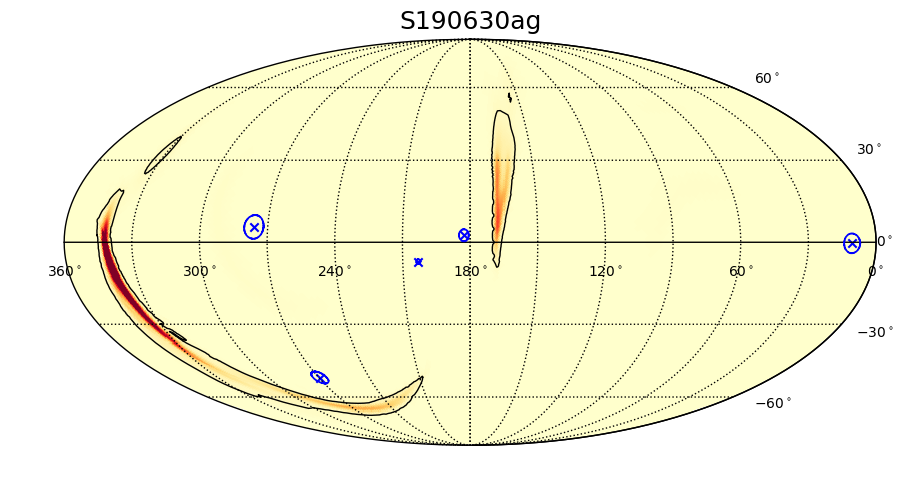} &
    \includegraphics[width=0.3\textwidth]{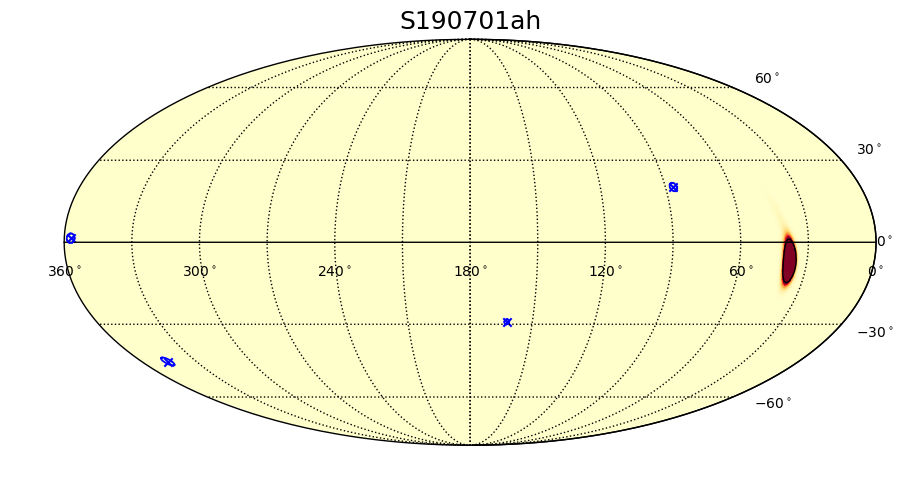} \\
    \includegraphics[width=0.3\textwidth]{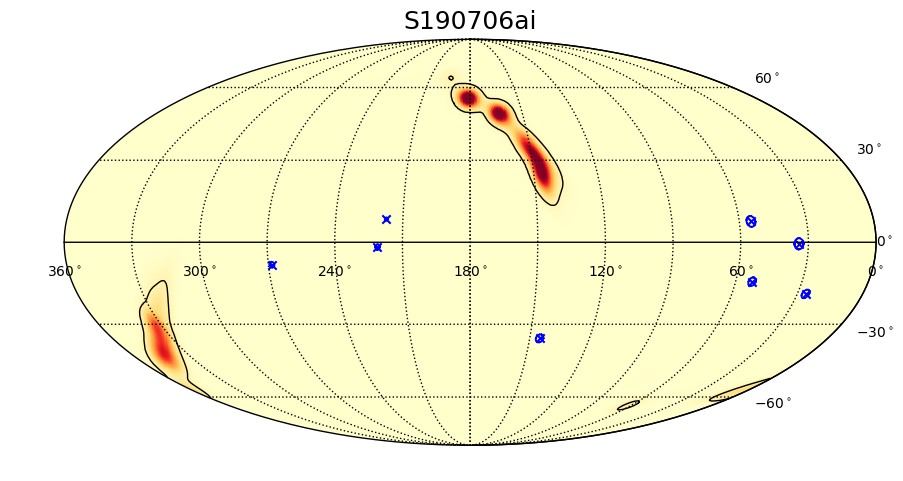} &
    \includegraphics[width=0.3\textwidth]{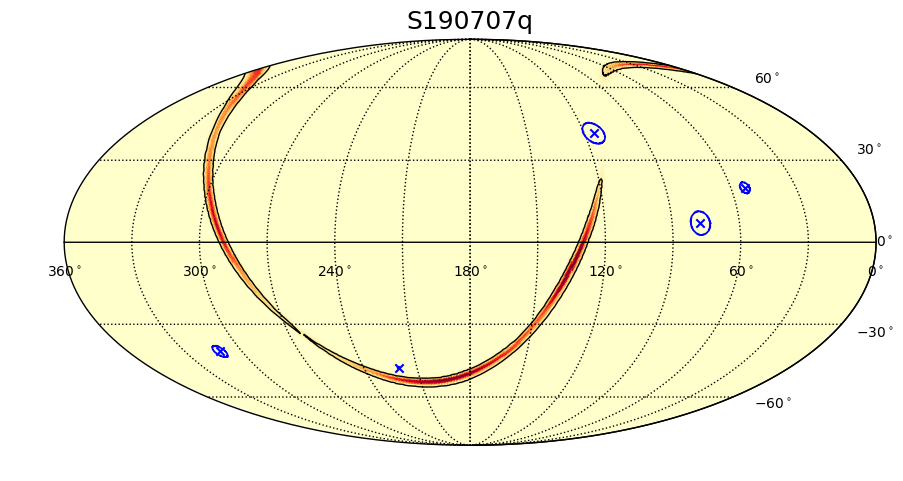} & 
    \includegraphics[width=0.3\textwidth]{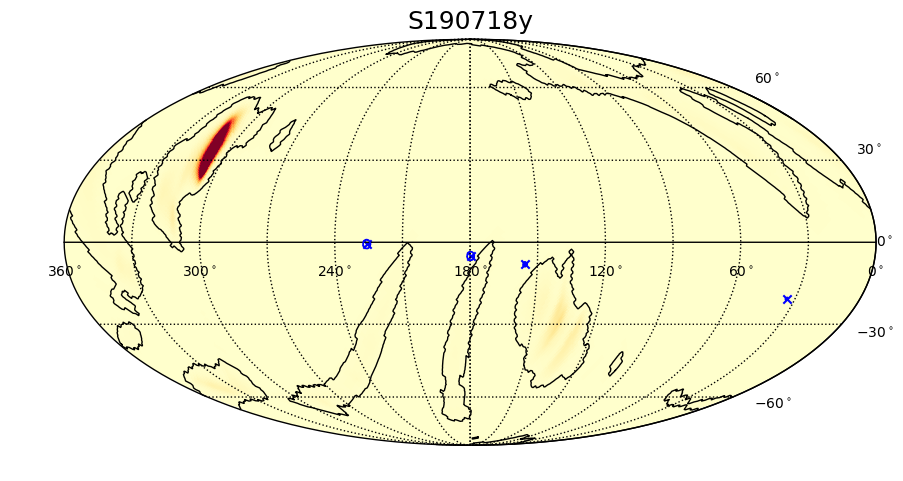}\\
    \includegraphics[width=0.3\textwidth]{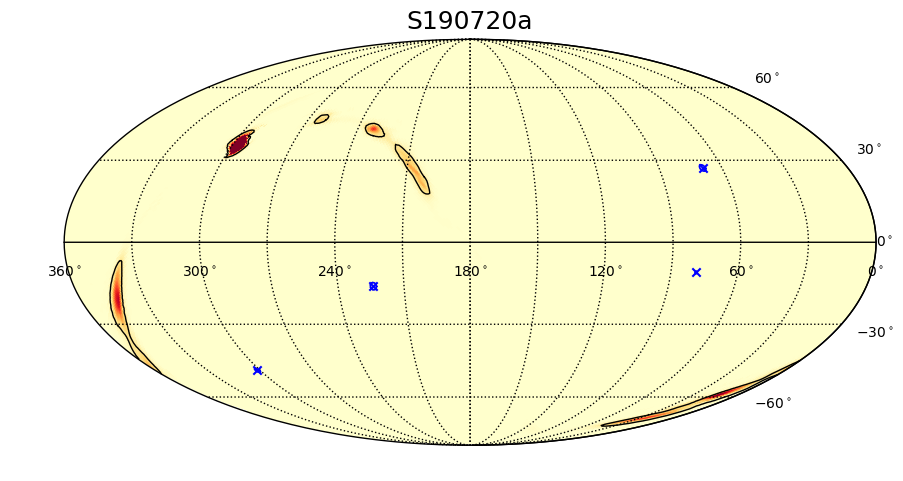} &
    \includegraphics[width=0.3\textwidth]{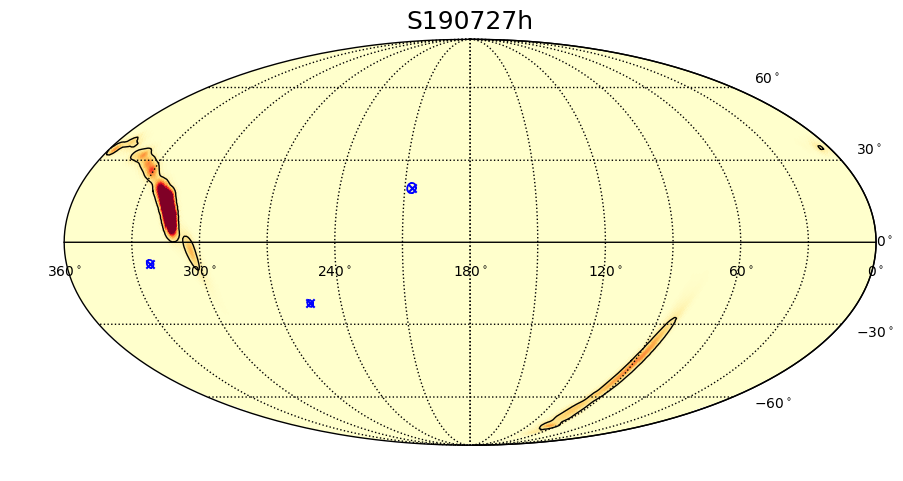} &
    \includegraphics[width=0.34\textwidth]{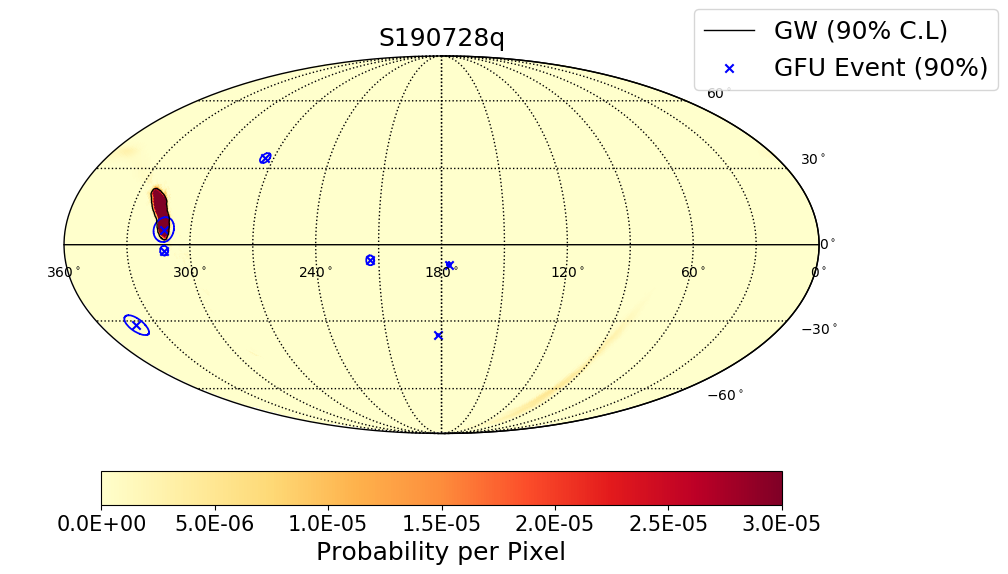} \\
    \multicolumn{3}{c}{\color{red}{{\Large{\textbf{IceCube Preliminary}}}}}\\
\end{tabular}
\end{center}
\caption{Joint skymaps for the first 33 detected GW events. Overlayed are the neutrinos within 1000 seconds of the GW trigger time. Gray contours around the blue crosses are 90\% containment angular errors for the neutrinos and the black contours are 90\% containment for the GW.}
\end{figure}

\newpage

\bibliographystyle{ICRC}
\bibliography{references}

\providecommand{\href}[2]{#2}\begingroup\raggedright\begin{thebibliography}{10}

\bibitem{LIGO:2018mvr}
{\bf LIGO Scientific, Virgo} Collaboration, B.~P. Abbott et~al.,
  \href{http://arxiv.org/abs/1811.12907}{{\tt arXiv:1811.12907}}.

\bibitem{graceDB}
``\textbf{Grace DB}.''
  https://gracedb.ligo.org/search/?query=\&query\_type=S\&results\_format=S.
\newblock Accessed: 2019-07-21.

\bibitem{GBM:2017lvd}
{\bf LIGO Scientific} Collaboration, B.~P. Abbott et~al., {\em Astrophys. J.}
  {\bf 848} (2017) L12.

\bibitem{ANTARES:2017bia}
{\bf ANTARES, IceCube, Pierre Auger, LIGO Scientific, Virgo} Collaboration,
  A.~Albert et~al., {\em Astrophys. J.} {\bf 850} (2017) L35.

\bibitem{Monitor:2017mdv}
{\bf LIGO Scientific, Virgo, Fermi-GBM, INTEGRAL} Collaboration, B.~P. Abbott
  et~al., {\em Astrophys. J.} {\bf 848} (2017) L13.

\bibitem{Halzen:2002pg}
F.~Halzen and D.~Hooper, {\em Rept. Prog. Phys.} {\bf 65} (2002) 1025--1078.

\bibitem{Razzaque:2003uv}
S.~Razzaque, P.~Meszaros, and E.~Waxman, {\em Phys. Rev.} {\bf D68} (2003)
  083001.

\bibitem{Murase:2015xka}
K.~Murase, D.~Guetta, and M.~Ahlers, {\em Phys. Rev. Lett.} {\bf 116} (2016)
  071101.

\bibitem{Adrian-Martinez:2016xgn}
{\bf ANTARES, IceCube, LIGO Scientific, Virgo} Collaboration,
  S.~Adrian-Martinez et~al., {\em Phys. Rev.} {\bf D93} (2016) 122010.

\bibitem{Albert:2018jnn}
{\bf ANTARES, IceCube, LIGO, Virgo} Collaboration, A.~Albert et~al., {\em
  Astrophys. J.} {\bf 870} (2019) 134.

\bibitem{Kintscher:2016uqh}
{\bf IceCube} Collaboration, T.~Kintscher, {\em J. Phys. Conf. Ser.} {\bf 718}
  (2016) 062029.

\bibitem{healpix}
K.~M. {G{\'o}rski}, E.~{Hivon}, A.~J. {Banday}, B.~D. {Wandelt}, F.~K.
  {Hansen}, M.~{Reinecke}, and M.~{Bartelmann}, {\em Astrophys. J.} {\bf 622}
  (Apr., 2005) 759--771.

\bibitem{Baret:2011tk}
B.~Baret et~al., {\em Astropart. Phys.} {\bf 35} (2011) 1--7.

\bibitem{Aartsen:2016lmt}
{\bf IceCube} Collaboration, M.~G. Aartsen et~al., {\em Astropart. Phys.} {\bf
  92} (2017) 30--41.

\bibitem{llama:2019icrc}
{\bf IceCube} Collaboration, A.~Keivani et~al.,  \pos{PoS(ICRC2019)930} (these
  proceedings).

\bibitem{Kimura:2018vvz}
S.~S. Kimura, K.~Murase, I.~Bartos, K.~Ioka, I.~S. Heng, and P.~Meszaros, {\em
  Phys. Rev.} {\bf D98} (2018) 043020.

\bibitem{Fang:2017tla}
K.~Fang and B.~D. Metzger, {\em Astrophys. J.} {\bf 849} (2017) 153.
  [Astrophys. J.849,153(2017)].

\end{thebibliography}\endgroup
\end{document}